\documentclass[aps,prb,preprint,groupedaddress,citeautoscript,nopacs]{revtex4}

\usepackage{graphicx}
\usepackage{color}
\usepackage{amsfonts}
\setcitestyle{super}

\begin{document}

\title{Topological protection of bound states against the hybridization}

\author{Bohm-Jung Yang,$^{1}$ Mohammad Saeed Bahramy,$^{1}$ and Naoto Nagaosa$^{1,2,3}$}

\affiliation{$^1$ Correlated Electron Research Group (CERG), RIKEN-ASI, Wako, Saitama 351-0198, Japan}

\affiliation{$^2$ Department of Applied Physics, University of Tokyo, Tokyo 113-8656, Japan}

\affiliation{$^3$ Cross-Correlated Materials Research Group (CMRG), RIKEN-ASI, Wako, Saitama 351-0198, Japan}

\date{\today}

\begin{abstract}
Topological invariants are conventionally known to be responsible
for protection of extended states against disorder.
A prominent example is the presence
of topologically protected extended-states in two-dimensional (2D)
quantum Hall systems as well as on the surface of three-dimensional (3D)
topological insulators. Distinct from such cases, here we introduce
a new concept, that is, the topological protection of bound states against hybridization.
This situation is shown to be realizable in a 2D quantum Hall insulator put on a 3D trivial insulator.
In such a configuration, there exist topologically protected bound states,
localized along the normal direction of 2D plane, in spite of hybridization
with the continuum of extended states. The one-dimensional edge states are
also localized along the same direction as long as their energies are within
the band gap. This finding demonstrates the dual role of topological invariants,
as they can also protect bound states against hybridization in a continuum.
\end{abstract}

\maketitle

Bound states of electrons in solids offer many intriguing phenomena.
For instance, localized states are common in the presence of impurities, and donor
and acceptor levels are the main issue in semiconductor physics and
technology. Bound states can also appear in clean systems
and their existence attributes to topological origins in many cases.
Solitons in polyacetylene are associated with the midgap states
which emerge due to topological reason, and control the electric, magnetic and
optical properties of the system.~\cite{Jackiw,Su,Goldstone}
One dimensional (1D) conducting channels at the edge of
the quantum Hall and quantum spin Hall system are another example of
bound states in clean systems, which determines the low energy transport phenomena.~\cite{Haldane,SHI_Murakami1,SHI_Murakami2,SHI_Zhang1,SHI_Zhang2}
Bulk-edge correspondence guarantees the existence of these edge channels,
i.e., they are protected by the gap and the nontrivial topology
of the bulk states.~\cite{BBC_Mong,BBC_Essin}
A recent remarkable advance in the study of
three dimensional topological insulators, which support the helical
Dirac fermion on the surface of the sample, provides another example of
bound states belonging to the same category.~\cite{TI_Kane,TI_Qi1,TI_Hasan,TI_Qi2}

A common feature of all these cases is that the energy of the bound
state is {\it within} the energy gap of the extended states.
The extended states constitute the continuum of the density of states
even though the disorder potential makes the crystal momentum $k$
ill-defined. Usually the extent of the localized state is determined by the
inverse of the energy separation between the localized level and
the edge of the continuum.
In general, it is believed that no state can be localized within this continuum
due to the hybridization with the extended states, as discussed in detail in Supplementary Note 1, although
there are several proposals for creating a bound state in a continuum
by engineering the effective coupling between the discrete level
and the continuum states.~\cite{BIC_Miyamoto, BIC_Longhi1, BIC_Longhi2}
This fact is the
essential reason for the existence of the mobility edge $E_c$ in disordered
systems.~\cite{MottDavis} Namely, there is a critical energy $E_c$ separating the extended and localized states.

In this paper, we study the electronic states of the
2D quantum Hall insulator (QHI)
on the substrate of a topologically trivial 3D normal insulator (NI).
One remarkable discovery through this study is that there are bound states even in the continuum
of the Bloch states of the substrate.
The hybridization with the Bloch states does not destroy the localization of some bulk states of the QHI
along the $z$-direction perpendicular to the surface. This is the first
example for the existence of bound states within the continuum of
extended states, which are protected by the topology.

\subsection*{Results}
{\bf Structure of the system.}
A schematic diagram describing the structure of the system
is shown in Fig.~\ref{fig:heterostructure}a.
We first consider a 2D QHI supporting 1D chiral edge states (CES) on its boundary and
a fully gapped 3D NI separately and then put the QHI on the top surface of the 3D NI.
As shown in Fig.~\ref{fig:heterostructure}b-c,
the typical band structure of the coupled system
is composed of three inequivalent parts, i.e.,
the 3D continuum states from the 3D NI and the 2D bulk and 1D CES from the QHI.
There are two main questions to address here. One is whether the 2D bulk states of the QHI
are exponentially localized near the top surface layer or they are spread out along the $z$-direction
when the 2D QHI and 3D NI are strongly hybridized, especially,
when the 2D QHI bands are fully buried in the continuum of the 3D NI bands.
The other is what the spatial distribution of the 1D CES is along the $z$-direction in this case.
These two issues are closely related because if the 2D states are extended and $\textit{diluted}$
along the $z$-direction, it is natural to expect the CES to be extended and diluted, as well.
As for the fate of 1D chiral states coupled to a continuum,
it is reminiscent of a recent theoretical work demonstrating the stability of the helical surface states
of topological insulators against the hybridization with additional non-topological bulk metallic states.~\cite{Bergman}
However, it is worth noting that the stability of 1D chiral states in our system is essentially a new issue
because when the 2D QHI is coupled to a 3D substrate, even the stability of the 2D QHI bulk state itself supporting the boundary chiral fermions
is not guaranteed.

To understand the effects of the hybridization between the 2D QHI and the 3D substrate,
we first study numerically a hexagonal lattice system composed
of stacked 2D honeycomb lattices as shown in Fig.~\ref{fig:heterostructure}d.
Here we treat the 3D NI as a system composed of coupled 2D layers
stacked along the $z$-direction, where each layer is labeled with the index $\ell_{z}=1,...,N_{z}$.
The top layer labeled with $\ell_{z}=0$ represents a 2D QHI, which is coupled to
the 3D NI through the inter-layer hopping transfer between nearest neighbors.
The detailed description of the lattice Hamiltonian is given in the Methods section.

{\bf Localized state in a continnum.}
In Fig.~\ref{fig:profile}, we plot the energy dispersion along the $k_{x}$-axis
and the localization profile of the system.
From  Fig.~\ref{fig:profile}a to Fig.~\ref{fig:profile}c, the energy gap of the 3D NI is gradually decreased, while
the other parameters of the QHI are fixed.
Also, layer-resolved wave function amplitudes are shown in each panel.
When the QHI bands are fully separated from the continuum as in Fig.~\ref{fig:profile}a,
the wave functions
of the QHI are almost completely localized in the first two layers.
On the other hand, Fig.~\ref{fig:profile}b shows that when the QHI and NI bands overlap in a finite interval,
the parts separated from (touching) the continuum are localized (extended).
However, surprisingly, the bound states still exist
even when the QHI bands are fully buried in the continuum as shown in Fig.~\ref{fig:profile}c.
In Fig.~\ref{fig:profile}c, there is a window near the corner of the Brillouin zone, marked with dotted circles, in which
bound states exist within the continuum states.
Especially, in this case, the states at this corner of the Brillouin zone
are completely localized, therefore their wave function amplitudes
become zero within the accuracy of our numerics for $\ell_{z}\geq5$.
In Fig.~\ref{fig:profile}g-h, we plot
the $\ell_{z}$ dependence of the squared wave function amplitude of the localized states
at the two corners of the hexagonal Brillouin zone.
To compare the wave function spreading at the two corners on the $k_{x}$-axis,
we have changed the strength of the inter-layer coupling between the top and
its neighboring layers relative to the inter-layer coupling inside the 3D NI, which is defined as $p$.
For given $p$ and the momentum, we have picked one state, among the eigenstates,
which has the highest wave function amplitude on the top layer.
When the top layer is decoupled from the rest of the system, that is, when $p=0$,
there are completely localized states at both corners. However, when $p$
increases to $p=0.5$ and finally to $p=1.0$, the state localized at the left corner
spreads rapidly along the $z$-direction (Fig.~\ref{fig:profile}h)
while the localized state on the right corner survives without any spreading of the wave function amplitude
(Fig.~\ref{fig:profile}g).
The distribution of localized states in the Brillouin zone is reflected in Fig.~\ref{fig:profile}d-f
in which the squared wave function amplitudes of the localized states
on the top layer ($\ell_{z}=0$) are plotted.

{\bf Topological stability of the localized state in a continuum.}
In fact, remarkably,
the bound state in a continuum has topological stability. Namely, the bound
state always exists in the Brillouin zone and cannot be removed by applying small perturbations.
It is to be noted that, as shown in Fig.~\ref{fig:profile}, the lattice model we have considered
has a 3-fold rotational symmetry ($C_{3}$) while all the other point group symmetries
are explicitly broken. To rule out the symmetry protection of the localized state
at the corner of the Brillouin zone, we have introduced a term
breaking $C_{3}$ symmetry and confirmed that
the localized state survives even when all the point group symmetries of the system
are completely broken.
In addition, the stability of such localized bound states is further supported by
additional numerical data demonstrating that localized bound states always exist irrespective of the lattice structure
and the nature of the inter-layer coupling between
the 2D QHI and 3D NI. The detailed information
about the additional numerical results is provided in Supplementary Figure S1 and Supplementary Note 2.

The topological stability of the localized state can be understood in the following way.
The minimal Hamiltonian for the 2D QHI and 3D NI, which describes one conduction
band and one valence band with a finite gap between them, can be written as
$H_{\text{QHI}}(\vec{k}_{\perp})=\vec{h}\cdot\vec{\tau}$ and
$H_{\text{NI}}(\vec{k}_{\perp},k_{z})=H_{0}+\vec{H}\cdot\vec{\tau}$
where $\vec{\tau}=(\tau_{x},\tau_{y},\tau_{z})$ are Pauli matrices.
Here $\vec{h}$ is a function of the in-plane momentum $\vec{k}_{\perp}=(k_{x},k_{y})$
while $H_{0}$ and $\vec{H}$ depend on the three momenta ($k_{x}$, $k_{y}$, $k_{z}$).
Considering the coupling between the 2D QHI and 3D NI, in general,
the retarded Green's function for the effective 2D QHI can be written as
\begin{equation}\label{eqn:retardedG}
G_{R}^{-1}(\Omega)=\Omega-\vec{h}\cdot\vec{\tau}
+i\frac{\Gamma}{2}\Big[1+a_{x}\tau_{x}+a_{y}\tau_{y}+a_{z}\tau_{z}\Big],
\end{equation}
where $\Gamma\neq0$ when $\Omega$ is within the 3D NI bands and
$\Gamma=0$ otherwise.
In Eq.~(\ref{eqn:retardedG}), the unit vector $\vec{a}=\pm\frac{\vec{H}}{|\vec{H}|}$,
in which $+$ sign ($-$ sign) corresponds to the case of $\Omega$ lying within the 3D bulk conduction (valence) band.
It is to be noted that here $\vec{a}$ and $\vec{H}$ depend on $(\Omega,k_{x},k_{y})$.
The $k_{z}$ dependence of $\vec{H}$ is replaced by the $\Omega$ dependence through
the $k_{z}$ integration in the self-energy correction.
Also the minor correction from the real part of the self-energy is neglected, however it
does not affect the generality of the conclusion.
The detailed derivation procedures to obtain $G_{R}$ can be found in Supplementary Note 3.
Then $\text{Det}G^{-1}_{R}=\Omega^{2}-h^{2}+i\Gamma(\Omega+\vec{a}\cdot\vec{h})=0$ immediately leads to
$\Omega=-\vec{a}\cdot\vec{h}=\pm |\vec{h}|$.
Namely, when the unit vector $\vec{a}$ is parallel (anti-parallel) to $\vec{h}$,
a Delta function singularity of $G_{R}$ appears at $\Omega=-|\vec{h}|$ ($\Omega=|\vec{h}|$),
which corresponds to the bound state and the localization along the $z$-direction.

Let us discuss about the physical meaning of the condition $\vec{a}=\pm\hat{h}\equiv\pm\frac{\vec{h}}{|\vec{h}|}$.
In the case of the QHI with the Hamiltonian $H_{\text{QHI}}=\vec{h}(\vec{k}_{\perp})\cdot\vec{\tau}$,
the unit vector $\hat{h}$ forms a skyrmion configuration
in the 2D momentum space and the wrapping
number (or the skyrmion number) of $\hat{h}$ over the first Brillouin zone
is equal to the Chern number of the QHI.~\cite{Qi_skyrmion}
On the other hand, since $\vec{H}$ stems from the Hamiltonian of the 3D NI, generally,
$\vec{a}=\vec{H}(\Omega,\vec{k}_{\perp})/|\vec{H}|$ cannot have nonzero
skyrmion number.
Namely, the unit vector $\vec{a}$ describes a
topologically trivial configuration.
Figure~\ref{fig:skyrmion}
shows the orientations of $\hat{h}$ and $\hat{a}$,
in which $\hat{h}$ forms a skyrmion configuration with the unit Pontryagin number.
It is to be noted that there are two special points, one at the center and the other at the boundary of the circle,
where $\hat{h}=-\hat{a}$ or $\hat{h}=\hat{a}$ is satisfied, respectively.
One of the two special points satisfies the condition
for bulk localization, either in the
conduction band or in the valence band.
Therefore there should be at least one localized 2D bulk state
in each of the conduction and valence bands.
The key ingredient leading to the topological protection of the localized bound state is that
the points satisfying $\hat{h}=\pm\hat{a}$ cannot be removed
as long as $\hat{h}$ and $\hat{a}$ describe topologically inequivalent configurations.
In general, the number of completely localized states buried in a continuum band is given
by $C_{\hat{h}}-C_{\hat{a}}$ where $C_{\hat{h}}$ is the skyrmion number (or Chern number)
covered by the unit vector $\hat{h}$.

{\bf Localization of 1D chiral edge states.}
Now we turn to the localization problem of the 1D CES.
To observe the 1D CES numerically,
we have introduced boundaries
with zigzag-type edges on each honeycomb layer.
It is confirmed that in all three cases corresponding to Fig.~\ref{fig:profile}a-c,
the 1D CES exists and is localized on the top few layers. Especially, as shown in Fig.~\ref{fig:profile}c and~\ref{fig:chiraledge},
it is valid even when the QHI bands are fully buried in the continuum, hence a part of the 1D CES is hybridized
with the continuum states. However,
the hybridization with
the continuum states does not simply destroy the 1D CES but instead the 1D CES
is pushed away from the overlapping region into the bulk gap, which is consistent with
the recent theory proposed by Bergman and Rafael~\cite{Bergman}.
The crucial point is that
the existence of in-gap states is solely determined by the topology of
$H_{\text{QHI}}$.
Moreover, once the CES exists, the fact that it is in the gap guarantees its localization
along the $z$-direction even if QHI bands are coupled to the continuum states.
The detailed proof is given in Supplementary Note 4.
Therefore as long as the gap between the conduction and valence
bands of the QHI is not fully buried in the 3D continuum,
the 1D CES should exist and be localized.

{\bf Extension to general cases.}
The topological protection of the localized bound states against hybridization
is a phenomenon that is valid in more general situations beyond the simple two-band model description
considered up to now.
Here we provide additional numerical data supporting the generality of
our theory.
To construct general lattice models, we have modified the in-plane dispersion
in the Hamiltonian of the 3D NI to lift its similarity to the Hamiltonian of the 2D QHI.
The detailed structure of the modified Hamiltonian describing the 3D NI is shown in the Methods section.

Let us discuss about the extension to multi-band systems. For simplicity,
we construct 2D four-band models attached to the two-band 3D NI.
To distinguish the different Chern number distribution between the four bands on the top layer
of the heterostructure,
we use the following notation of $\textbf{C}\equiv[C_{1},C_{2},C_{3},C_{4}]$
where $C_{i}$ indicates the Chern number of the $i$-th band.
Here the four bands are aligned in decreasing order of energy
from the band 1 with the highest energy to the band 4 with the lowest energy.

We first consider the case of $\textbf{C}=[0,0,-1,+1]$.
Here the two high (low) energy bands are buried in the conduction (valence) band continuum of the 3D NI.
From the 2D distribution of the states localized along the $z$-direction
as shown in the two right panels of Fig.~\ref{fig:4band}a, we can clearly see
the different degrees of localization between the topologically trivial bands and non-trivial bands buried in a continuum.
This is also supported by the layer-resolved wave function amplitudes.
Interestingly, even when the two 2D bands with opposite Chern numbers are buried in a single
continuum band, so that the gap between the 2D bands is filled with continuum states,
each of the 2D bands supports completely localized bulk states separately.
This shows the fact that the number of localized 2D bulk states is solely determined
by the Chern number difference between the 2D bulk band and the effective 2D band of
the 3D NI, which is consistent with the conclusion based on the Green's function analysis given above.
On the other hand, as for the localization of the 1D CES, the existence of the band gap between the 2D bands supporting
the CES is crucial.
From the numerical analysis of the edge spectrum,
it is confirmed that there is no 1D CES localized along the z-direction in this case.
We also have considered other model systems such as
$\textbf{C}=[-1,0,+1,0]$ (Fig.~\ref{fig:4band}b),
$\textbf{C}=[0,-1,1,0]$ (Fig.~\ref{fig:4band}c),
$\textbf{C}=[-1,-1,+1,+1]$ (Fig.~\ref{fig:4band}d),
$\textbf{C}=[-1,+1,+1,-1]$ (Fig.~\ref{fig:4band}e) cases.
All of them support the same conclusion that the localization
of the 2D bulk state is entirely determined by the Chern number of the
2D band buried in the continuum states of the topologically trivial 3D NI. However, the existence of the band gap
is required for the localization property of the 1D CES
in addition to the topological property of the 2D bulk states supporting the 1D CES.

Next we describe the localization property of the 2D QHI with higher Chern number,
in particular, when the Chern number of the conduction (valence) band is equal to -2 (+2).
In this case, we expect that there should be at least
two localized 2D bulk states in each of the valence or conduction band when
each band is completely buried in a continuum.
As shown in Fig.~\ref{fig:C2}, we can clearly see that there are several
localized states which are well-separated in the momentum space.
In particular, along the boundary of the Brillouin zone, it is numerically
confirmed that there are three completely localized states. Among them,
two states are related by the reflection symmetry but the other
one is not. Therefore there are at least two inequivalent localized states in the Brillouin
zone, consistent with the prediction based on the consideration of topological invariants.

Finally, let us consider the localization property of
the 2D quantum spin Hall insulator on a substrate.
Since the QSHI can be considered as a superposition of two QHI systems possessing opposite Chern numbers,~\cite{TI_Kane}
the topological protection of the bulk and edge states is expected to be observed in the QSHI heterostructure.
As for the topological property of the QSHI, it is important to clarify
whether the $z$-component of the total spin, i.e., $S_{z}$ is conserved or not.
When $S_{z}$ is conserved (not conserved), the topological invariant of the 2D system
is the spin Chern number (the $Z_{2}$ invariant).
In addition, we consider two different model Hamiltonians for the 3D NI.
One is the two-band model breaking the time-reversal symmetry and
the other is the four-band model preserving the time-reversal symmetry.
In the case of the two-band model on the honeycomb lattice, since there is only one available state
in each site, it can be considered as a spin-polarized system breaking the time-reversal symmetry.

As shown in Fig.~\ref{fig:QSH}a-b, when the 3D NI breaks
the time-reversal symmetry,
several localized bound states are found independent
of the presence (Fig.~\ref{fig:QSH}a) or absence (Fig.~\ref{fig:QSH}b) of the $S_{z}$ conservation.
On the other hand, when the total heterostructure maintains
the time-reversal symmetry, the $S_{z}$ conservation is crucial to obtain localized bound states.
When the $S_{z}$ is conserved, several completely localized
bound states appear as shown in Fig.~\ref{fig:QSH}c.
However, surprisingly, once the $S_{z}$ conservation is violated due
to the Rashba-type spin-mixing term,
all the 2D bulk states of the QSHI spread along the $z$-direction, as shown in Fig.~\ref{fig:QSH}d.
It is numerically confirmed that there is no localized bound state in the Brillouin zone.
This observation demonstrates the fundamental difference between the Chern number (or spin Chern number)
and the $Z_{2}$ invariant. In contrast to the case of the Chern number,
the nonzero $Z_{2}$ invariant cannot protect any localized bound state in a continuum state.

\subsection*{Discussion}

The existence of 2D bulk states localized along the $z$-direction in a continuum protected by topology
reveals the dual role played by the topological invariant.
Conventionally, the topological invariant has been considered as a quantity which protects
the {\it extended} bulk states against the localization due to disorder.
These extended bulk states protected by topology guarantee the existence of the delocalized states on the sample boundary
even in the presence of disorder,
which gives rise to the profound relationship between the bulk topological invariant
and the extended boundary states, dubbed the bulk-boundary correspondence.
For example, in integer quantum Hall systems, the Chern number can be
formulated in terms of the number of chiral edge modes.~\cite{QHI_Laughlin,QHI_Halperin, QHI_Thouless} Also,
the $Z_{2}$ invariants of 2D (3D) time-reversal invariant systems
determine the parity of the number of helical edge modes (Dirac cones) at the boundary.~\cite{TI_FuKane}
These boundary states within the bulk band gap smoothly connect the bulk valence and conduction bands,
which can be deformed and fully merged into the bulk bands only when the corresponding bulk topological invariant
changes through the bulk band gap-closing.
Namely, the pair annihilation of the topological invariants is the only mechanism to eliminate the extended states.

On the other hand, when the topological insulator is coupled to
a substrate material, the topological invariant also guarantees the existence of
some bulk states {\it localized} along the $z$-direction, which are effectively decoupled from the continuum
states of the substrate. In the case of the 2D QHI on a topologically trivial
3D NI, the Chern number of the QHI determines
the minimum number of localized 2D QHI bulk states buried in the continuum states of the 3D NI.
The essential ingredient leading to the localized 2D bulk states buried in a continuum
is the inequivalence of the Chern numbers between the 2D QHI and the effective 2D system
constituting the 3D NI.
Therefore if the topological property of the QHI on the top layer can be controlled
by tuning the bulk band gap of the 2D QHI, the completely localized states
can start to be hybridized with the continuum states of the 3D NI across the topological phase transition
and be spread through the whole sample in the end when the 2D bulk state on the top layer is deep inside of
the topologically trivial phase.

In addition, using the same framework, it is also possible to
understand the evolution of the energy band structure of the coupled 2D QHI systems stacked along the
vertical $z$-direction.
Here we assume that the inter-layer coupling is not so strong
to induce any band touching through the inter-layer hybridization.
If we distinguish the QHI on the top layer from the other part of the system,
the whole system can be considered as a heterostructure composed
of a 2D QHI layer and a 3D QHI substrate.
Since the top layer and the effective 2D layer of the 3D substrate
have the same Chern number, there can be no 2D bulk state that is localized
on the top layer, which is confirmed through the additional numerical study discussed in detail in
Supplementary Note 5 with figures shown in Supplementary Figure S2.
In this case, the 1D chiral edge states of the top QHI layer
should also be delocalized and spread along the vertical $z$-direction.
The extended 1D chiral edge states eventually participate in the formation of the boundary gapless states
of the 3D weak QHI,~\cite{3DQHI} which disperse along the two orthogonal directions,
one of which corresponds to the vertical $z$-direction.~\cite{3DQHI_leon}
To establish the general relationship between
the topological invariant of the 3D substrate and the localization property of the 2D QHI
is an interesting problem requiring additional studies.

Let us briefly discuss the influence of disorder.
As shown above, the bound states exponentially localized along $z$-direction
exist only at several discrete $k$-points.
Since the disorder scattering mixes the wavefunctions
at different $k$-points, it is expected that the bound states can be hybridized
with extended states and hence delocalized along $z$-direction.
Correspondingly, the delta-functional peak of the spectral function will turn into a broadened resonance.
The edge channels, on the other hand,
can be protected by the gap, and hence remain localized along $z$-direction.
However, these hand-waving arguments should be confirmed by more careful analysis, which is left for future studies.

We conclude with a discussion of candidate materials
realizing the topological protection of localization.
A recent LDA+U study has proposed that several pyrochlore
iridates can be anti-ferromagnetic insulators
with the all-in/all-out spin configuration.~\cite{Iridate}
In this magnetic state, the [111]-surface or any of its symmetry equivalents
supports a kagome-lattice plane with a non-coplanar spin ordering.
Since the spin chirality induced by non-coplanar spins
can give rise to a QHI on the surface~\cite{Kagome_QH}
while the bulk insulating state is topologically trivial,
pyrochlore iridates are an ideal platform
to realize intriguing localization phenomena protected by topology.

{\small \subsection*{Methods}
{\bf Lattice Hamiltonian of the stacked 2D honeycomb lattice system.}
In each honeycomb layer, in addition to the hopping amplitude $t$ between nearest-neighbor sites,
we take into account of two additional terms, i.e., the staggered chemical potential $\mu_{s}$ between
two sublattice sites in a unit cell and the imaginary hopping amplitude $\lambda_{\text{so}}$ between next nearest neighbor sites.
Explicitly, the Hamiltonian including $\mu_{s}$ and $\lambda_{\text{so}}$ can be written as
\begin{eqnarray}
H_{\mu_{s},\lambda_{\text{so}}}&=&-t\Big[(1+\cos k_{1}+\cos k_{2})\tau_{x}+(\sin k_{1}+\sin k_{2})\tau_{y}\Big]
\nonumber\\
&+&\Big[\mu_{s}-2\lambda_{\text{so}}(\sin k_{1} -\sin k_{2} -\sin(k_{1}-k_{2}))\Big]\tau_{z},
\nonumber
\end{eqnarray}
where $\tau_{x,y,z}$
are Pauli matrices indicating the two sublattice sites and $k_{1}=\frac{1}{2}k_{x}+\frac{\sqrt{3}}{2}k_{y}$,
$k_{2}=-\frac{1}{2}k_{x}+\frac{\sqrt{3}}{2}k_{y}$.
In this Hamiltonian, when $\mu_{s}<3\sqrt{3}\lambda_{\text{so}}$ ($\mu_{s}>3\sqrt{3}\lambda_{\text{so}}$),
the system is a QHI (a trivial insulator).~\cite{Haldane}
Here we take $H_{\mu_{s}=0,\lambda_{\text{so}}}$ as a model Hamiltonian
for QHI. Also we use $H_{\mu_{s},\lambda_{\text{so}}=0}$ as a basis
to construct a 3D NI by incorporating inter-layer hopping amplitudes.
Then the Hamiltonian for the coupled system can be written as
\begin{eqnarray}\label{eqn:latticeH}
H_{\text{coupled}}&=&\sum_{\textbf{k}_{\perp}}\sum_{\ell_{z}=0}^{N_{z}}\psi^{\dag}_{\ell_{z}}
H_{\mu_{s},\lambda_{\text{so}}}(\textbf{k}_{\perp};\ell_{z})\psi_{\ell_{z}}
\nonumber\\
&+&\sum_{\textbf{k}_{\perp}}\sum_{\ell_{z}=0}^{N_{z}-1}\Big\{\psi^{\dag}_{\ell_{z}}
(T_{0}+T_{z}\tau_{z})\psi_{\ell_{z}+1}+h.c.\Big\},
\nonumber
\end{eqnarray}
where $H_{\mu_{s},\lambda_{\text{so}}}(\textbf{k}_{\perp};\ell_{z}=0)=H_{\mu_{s}=0,\lambda_{\text{so}}}(\textbf{k}_{\perp})$
and $H_{\mu_{s},\lambda_{\text{so}}}(\textbf{k}_{\perp};\ell_{z}\neq 0)=H_{\mu_{s},\lambda_{\text{so}}=0}(\textbf{k}_{\perp})$.
For numerical computation, we choose $\lambda_{\text{so}}=1$, $T_{0}=3$, $T_{z}=0.3$.
$t=1$ ($t=3.5$) for $\ell_{z}\neq0$ ($\ell_{z}=0$). To control the energy gap
of 3D NI we use $\mu_{s}=20$ for Fig.~\ref{fig:profile} (a),
$\mu_{s}=12$ for Fig.~\ref{fig:profile} (b),
$\mu_{s}=7.5$ for Fig.~\ref{fig:profile} (c).

For the construction of general lattice models,
we replace the basis for the 3D NI by the following Hamiltonian,
\begin{equation}
H_{\text{NNN}}=\mu'\tau_{x}+t'\Big[\cos k_{1} + \cos k_{2} + \cos (k_{1}-k_{2})\Big]\tau_{z},
\end{equation}
where $t'$ is the hopping amplitude between next nearest neighbor sites.
Here the hybridization between two sublattice sites $\mu'$ determines
the size of the band gap. The Hamiltonian for a 3D NI can be obtained
by stacking $H_{\text{NNN}}$ along the $z$-direction.

{\bf Existence of 1D chiral edge state under hybridization.}
The influence of
the coupling to the 3D continuum states on the existence of
the 1D chiral edge states can be considered in the following way. We first consider a decoupled
2D QHI described by the Green's function $G_{\text{QHI}}^{(0)}(x,x';k_{y};\Omega)$ where $x$ ($k_{y}$)
is the spatial (momentum) coordinate and $\Omega$ is the energy.
Let $V(x_{0})$ be a local perturbation creating a boundary at $x=x_{0}$
parallel to the $y$ direction. Then the system with a boundary can be described
by the Green's function $G^{(0)}(x,x';\Omega)$
which satisfies the equation $G^{(0)}(x,x';\Omega)=G_{\text{QHI}}^{(0)}(x,x';\Omega)+G_{\text{QHI}}^{(0)}(x,x_{0};\Omega)VG^{(0)}(x_{0},x';\Omega)$
where the $k_{y}$ coordinate is omitted.~\cite{Niu}
After simple algebra, we can show that
$G^{(0)}(x_{0},x';\Omega)=\Big\{1/[1-G_{\text{QHI}}^{(0)}(x_{0},x_{0};\Omega)V]\Big\}G_{\text{QHI}}^{(0)}(x_{0},x';\Omega)$
which leads to the following condition to obtain the edge state
\begin{equation}\label{eqn:edgecondition}
\text{Det}[1-G_{\text{QHI}}^{(0)}(x_{0},x_{0};\Omega)V]=0.
\end{equation}
Now we turn on the coupling to the 3D continuum state
under which $G_{\text{QHI}}^{(0)}$ promotes to $G_{\text{QHI}}$
satisfying $G_{\text{QHI}}(\Omega+i0^{+})=G_{R}(\Omega)$, the retarded Green's function in Eq.~(\ref{eqn:retardedG}).
Considering the local potential $V(x_{0})$ again, the condition to obtain
an edge state can be derived simply by replacing $G_{\text{QHI}}^{(0)}$
by $G_{\text{QHI}}$ in Eq.~(\ref{eqn:edgecondition}).
The point is that when $\Omega$ is within the bulk gap of 3D substrate
$G_{\text{QHI}}=G_{\text{QHI}}^{(0)}$ because $G_{R}(\Omega)$
develops the imaginary part only when $\Omega$ is within the 3D continuum states.
Therefore the coupling to 3D continuum states does not affect
the existence of edge states.

Moreover, in the case of the zero energy edge state,
the condition to obtain the in-gap state can be rewritten in terms of
the thermal Green's function
at zero Matsubara frequency $\mathcal{G}^{-1}(i\omega=0)$.
Namely, $\text{Det}[1-\mathcal{G}(x_{0},x_{0};i\omega=0)V]=0$
determines the condition to obtain the zero-energy in-gap state.
It is interesting to note the consistency of this result with
the recent work proposing that the topological
invariant of general interacting systems
can be written in terms of the Green's function
at zero Matsubara frequency.~\cite{Wang}

\bibliographystyle{naturemag}


{\small \subsection*{Acknowledgements}
We thank Patrick A. Lee for his insightful comments.
We are grateful for support from the Japan Society for the Promotion of Science (JSPS) through the `Funding Program for World-Leading Innovative R\&D on Science and Technology (FIRST Program)}

{\small \subsection*{Author contributions}
B.-J.Y., M.S.B, and N.N. conceived the original ideas.
B.-J.Y. did the numerics.
B.-J.Y., M.S.B, and N.N. analyzed
the data and wrote the manuscript.
}




\

\begin{figure*}[t]
\centering
\includegraphics[width=16 cm]{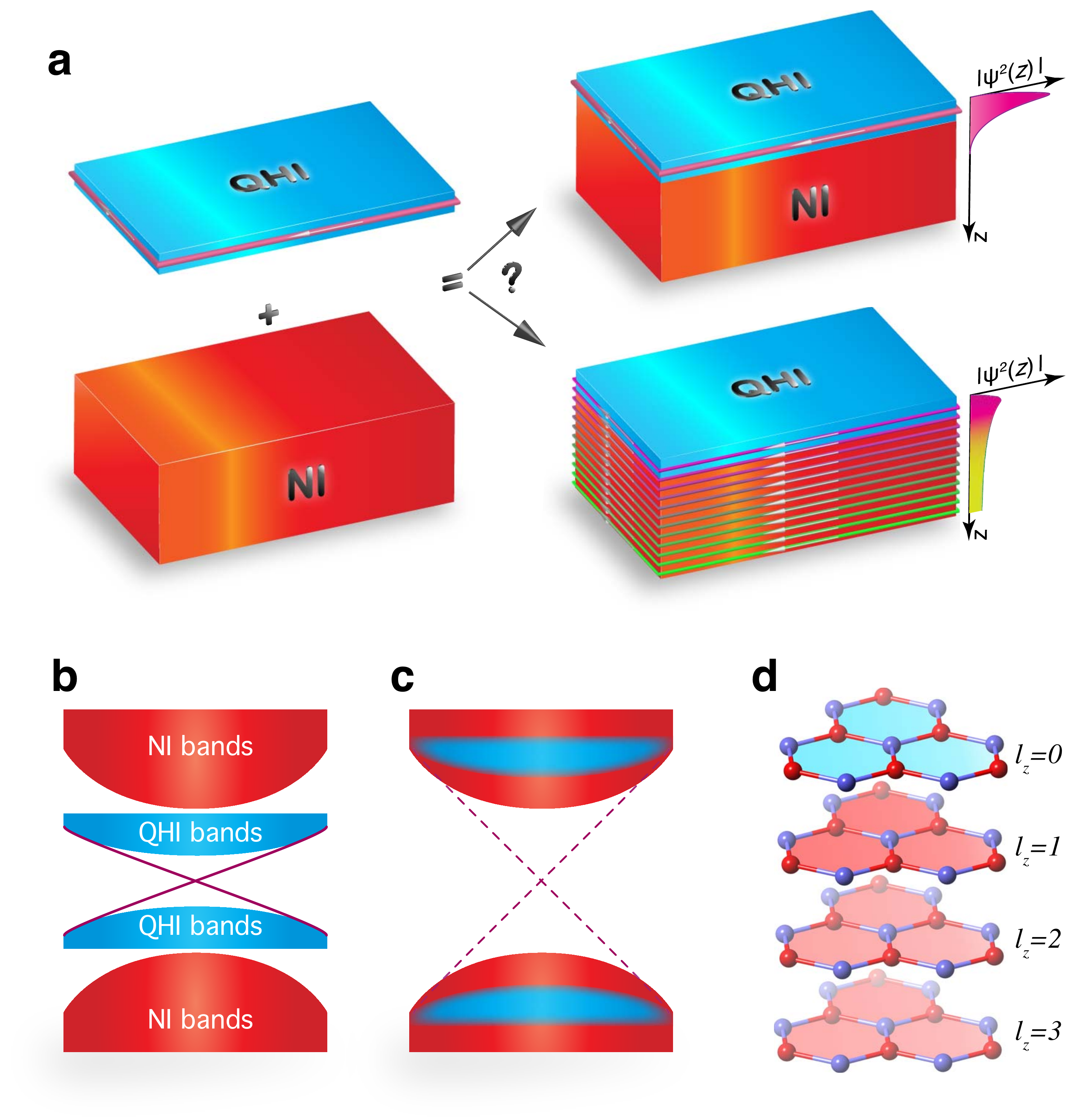}
\caption{
{\bf Geometry and energy band diagram of the heterostructure composed of two-dimensional (2D)
quantum Hall insulator (QHI) and three-dimensional (3D) normal insulator (NI).}
({\bf a}) A schematic diagram showing the structure of the system.
A 2D QHI with one-dimensional (1D) chiral edge state is attached on the top surface of a 3D NI.
One main question is whether the 1D chiral edge state is localized or extended along the $z$-direction
when 2D QHI and 3D NI are coupled.
The typical energy spectra of the full heterostructure in the momentum space are shown in ({\bf b}) and ({\bf c}).
The NI bands (QHI bands) are the states coming from NI (QHI)
and the dispersion in the gap corresponds to the (possible) 1D
edge channel.
({\bf d}) A hexagonal lattice model composed of 2D honeycomb lattices
stacked along the $z$-direction. The QHI on the top layer ($\ell_{z}=0$)
is coupled to the 3D NI composed of stacked 2D trivial insulator layers ($\ell_{z}\neq0$).
} \label{fig:heterostructure}
\end{figure*}

\begin{figure*}[t]
\centering
\includegraphics[width=16 cm]{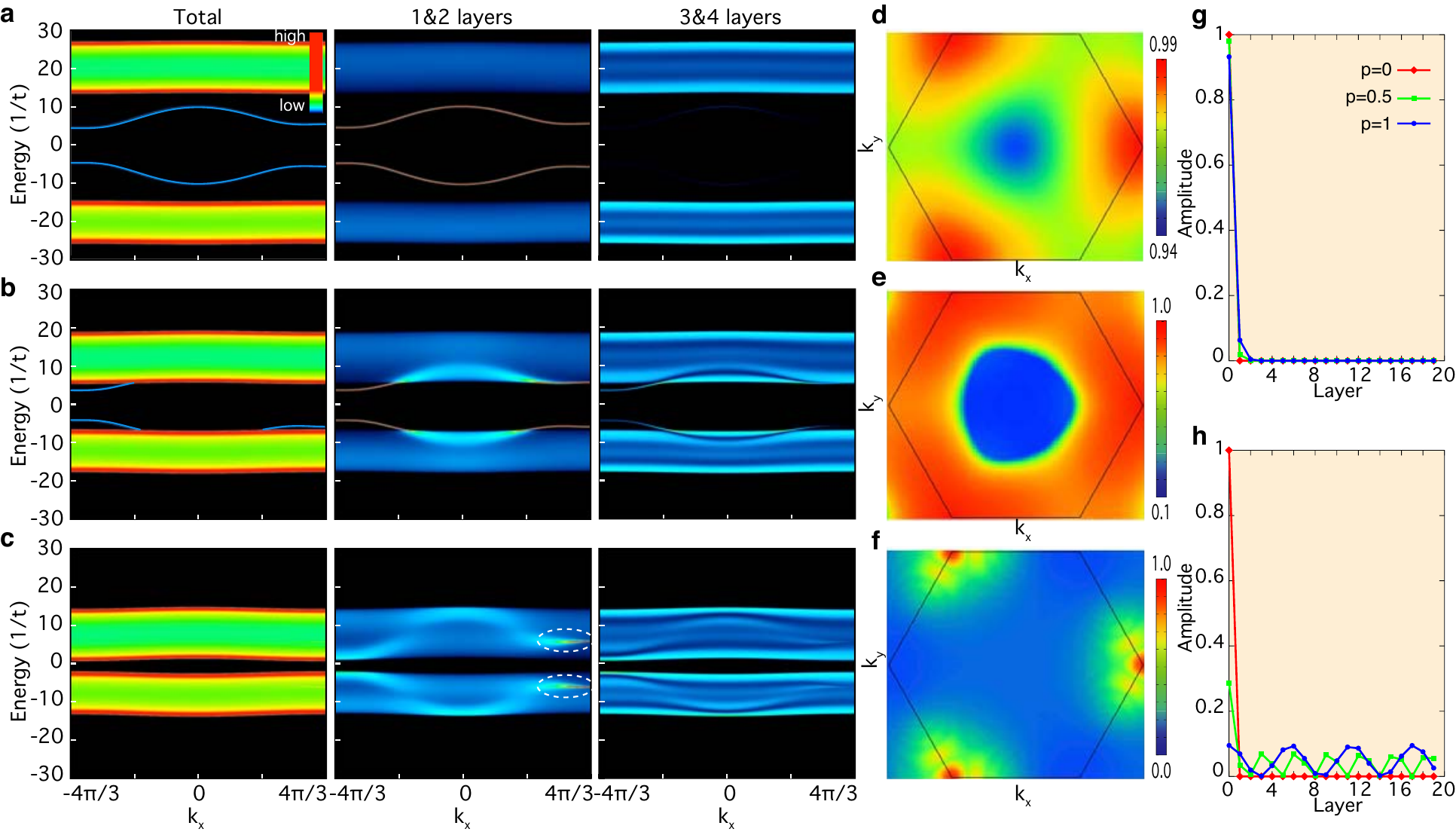}
\caption{
{\bf Energy dispersion and distribution of localized states of two dimensional (2D)
quantum Hall insulator (QHI) bulk bands in a continuum.}
Energy dispersion and the distribution of localized 2D bulk states
of the coupled system composed of honeycomb layers stacked along the $z$-direction.
Bulk band structure of the system along $k_{y}=0$ is shown
in ({\bf a}), ({\bf b}), and ({\bf c}) where the normal insulator (NI) bands and QHI bands
are not touching, partially ovarlapped, and fully overlapped, respectively.
From ({\bf a}) to ({\bf c}), the energy gap of the NI band is progressively reduced while
the spectrum of the QHI band remains the same.
In each panel, the layer-resolved wave function amplitudes of
$|\psi_{n,\textbf{k}_{\perp}}(\ell_{z})|^{2}$ are shown.
Here $\psi_{n,\textbf{k}_{\perp}}$
is the n-th eigenfunction at the momentum $\textbf{k}_{\perp}$.
The localized bound states are almost completely confined within the first two layers.
The dotted circles on the second figure from the left in the panel ({\bf c})
indicate the localized bound states existing within the continuum states.
In ({\bf d}), ({\bf e}), ({\bf f}), we have picked up one state among all the eigenstates at each $\textbf{k}_{\perp}$, which has the maximum value
of $|\psi_{n,\textbf{k}_{\perp}}(\ell_{z}=0)|^{2}$,
and then the resulting squared amplitude of $|\psi_{n,\textbf{k}_{\perp}}(\ell_{z}=0)|^{2}$
is plotted over the entire Brillouin zone
for systems corresponding to ({\bf a}), ({\bf b}), ({\bf c}), respectively.
Finally, in ({\bf g}) and ({\bf h}), the $\ell_{z}$-dependence
of the squared wave function amplitudes $|\psi_{n,\textbf{k}_{\perp}}(\ell_{z})|^{2}$ of the localized state
is plotted for $\textbf{k}_{\perp}=(\frac{4\pi}{3},0)$ ({\bf g}) and
for $\textbf{k}_{\perp}=(-\frac{4\pi}{3},0)$ ({\bf h}).
Here we have varied the strength of the inter-layer coupling between the top and its neighboring layers
relative to the inter-layer coupling inside the 3D NI, which is defined as $p$.
For fixed $p$ and $\textbf{k}_{\perp}$, we have picked the eigenfunction
which has the highest amplitude on the top layer and plotted its $\ell_{z}$ dependence.
} \label{fig:profile}
\end{figure*}

\begin{figure*}[t]
\centering
\includegraphics[width=12 cm]{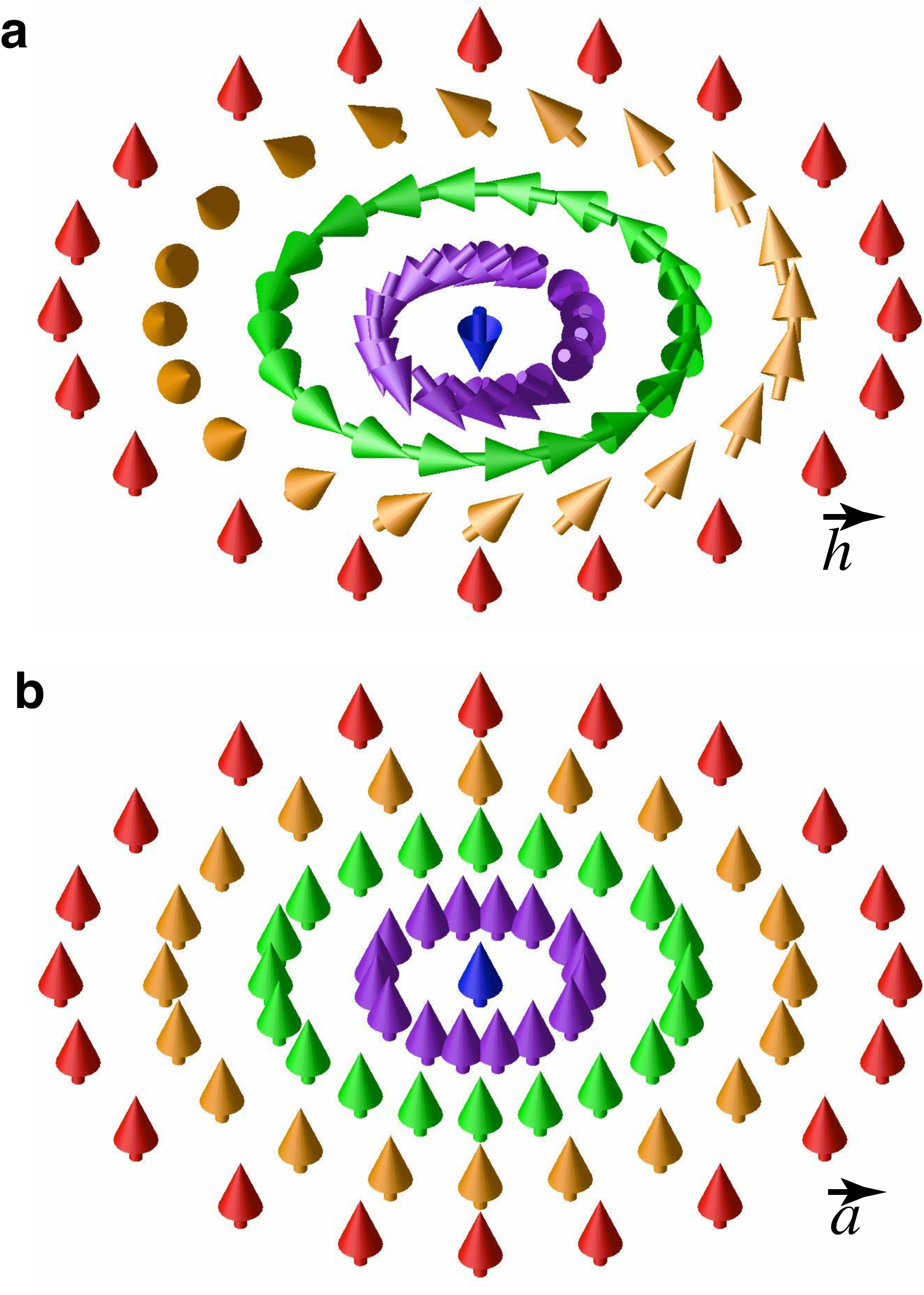}
\caption{{\bf Topological origin of bulk localization.}
Schematic figures showing the topological structure of
the unit vectors $\vec{h}$ and $\vec{a}$ in a 2D space.
$\vec{h}$ forms a skyrmion configuration with the unit Pontryagin number while $\vec{a}$
makes a topologically trivial structure. $\vec{h}$ and $\vec{a}$ are parallel (anti-parallel)
at the boundary (at the center).
} \label{fig:skyrmion}
\end{figure*}

\begin{figure*}[t]
\centering
\includegraphics[width=16 cm]{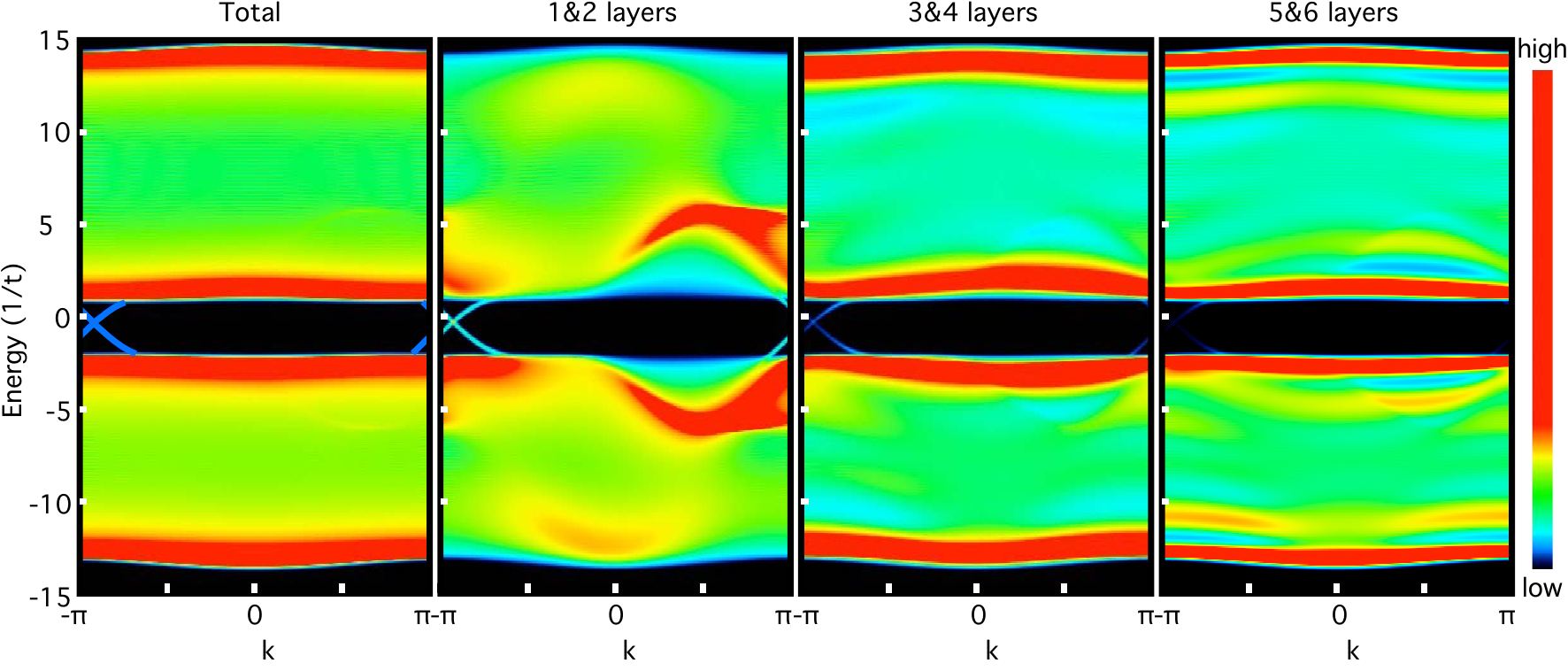}
\caption{{\bf Localization of one-dimensional (1D) chiral edge state.}
Energy dispersion and layer-resolved wave function amplitudes
of the finite size system corresponding to Fig.~\ref{fig:profile}c.
To obtain the 1D chiral edge spectrum, we introduce two parallel zigzag edges
in each honeycomb layer and
stack the corresponding strip structures along the $z$-axis.
Then the resulting system possesses two parallel surfaces whose surface normal direction is perpendicular
to the $z$-axis.
} \label{fig:chiraledge}
\end{figure*}

\begin{figure*}[t]
\centering
\includegraphics[width=16 cm]{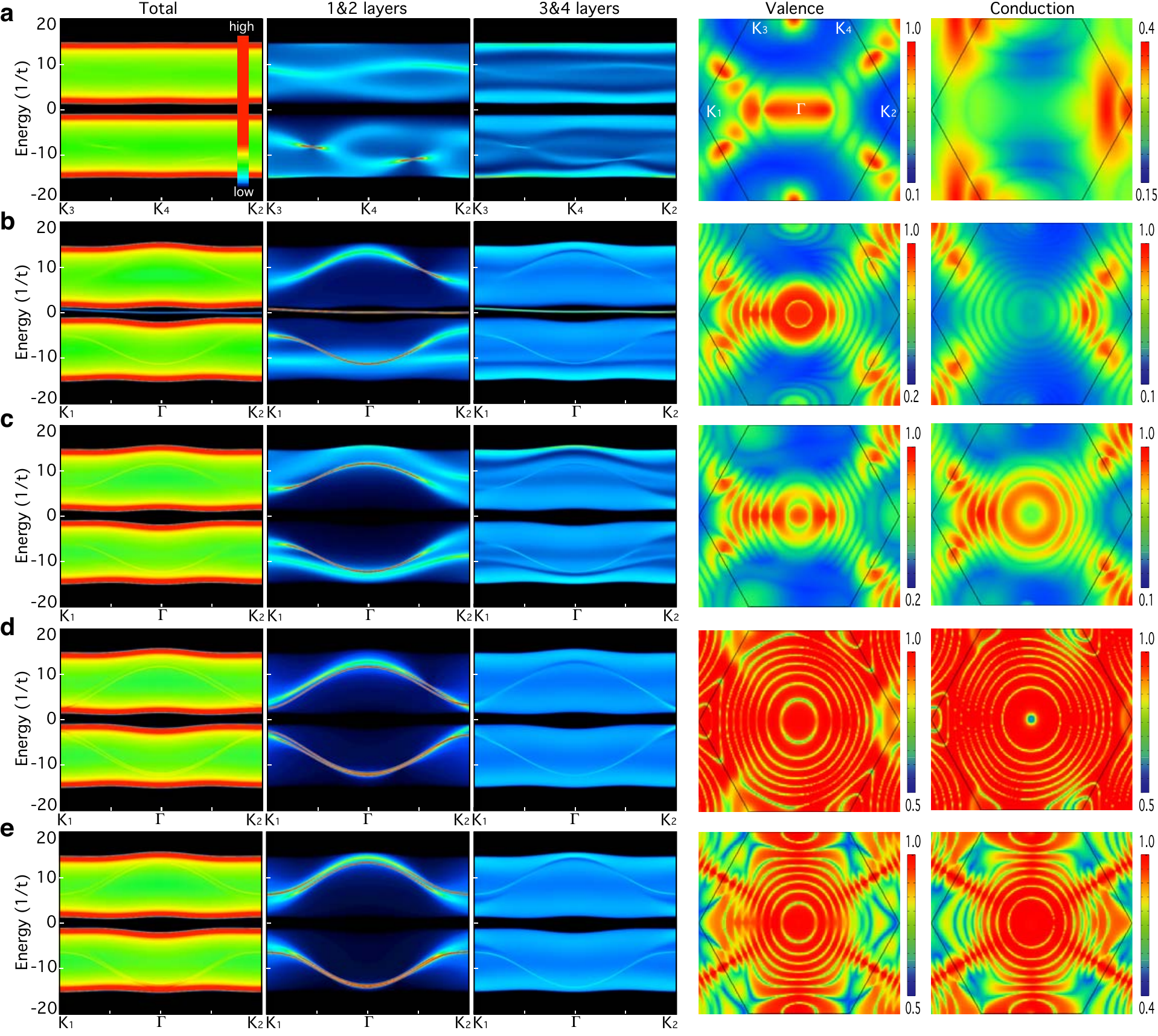}
\caption{{\bf Energy dispersion and distribution of localized states of two-dimensional (2D) four-band
quantum Hall insulator (QHI) bulk bands in a continuum
of the three-dimensional (3D) normal insulator (NI).}
Energy dispersion and localization profile
of the coupled system composed of honeycomb layers stacked along the $z$-direction.
For each model system, we plot the layer-resolved wave function amplitudes
along a particular direction in the momentum space and the 2D distribution of states localized along
the $z$-direction.
In each model, the Chern number distribution of the four bands on the top surface layer
is indicated by a vector $\textbf{C}=[n_{1},n_{2},n_{3},n_{4}]$.
In all cases except the case ({\bf b}), the two high (low) energy bands with the Chern number
$n_{1}$, $n_{2}$ ($n_{3}$, $n_{4}$) are buried in the conduction (valence) band of the 3D NI.
In the case of ({\bf b}), the second band with the Chern number $n_{2}=0$ is in the gap.
({\bf a}) $\textbf{C}=[0,0,-1,+1]$.
({\bf b}) $\textbf{C}=[-1,0,+1,0]$.
({\bf c}) $\textbf{C}=[0,-1,+1,0]$.
({\bf d}) $\textbf{C}=[-1,-1,+1,+1]$.
({\bf e}) $\textbf{C}=[-1,+1,+1,-1]$.
In all cases, only the 2D bands with nonzero Chern numbers support bound states
localized along the $z$-direction.
} \label{fig:4band}
\end{figure*}

\begin{figure*}[t]
\centering
\includegraphics[width=16 cm]{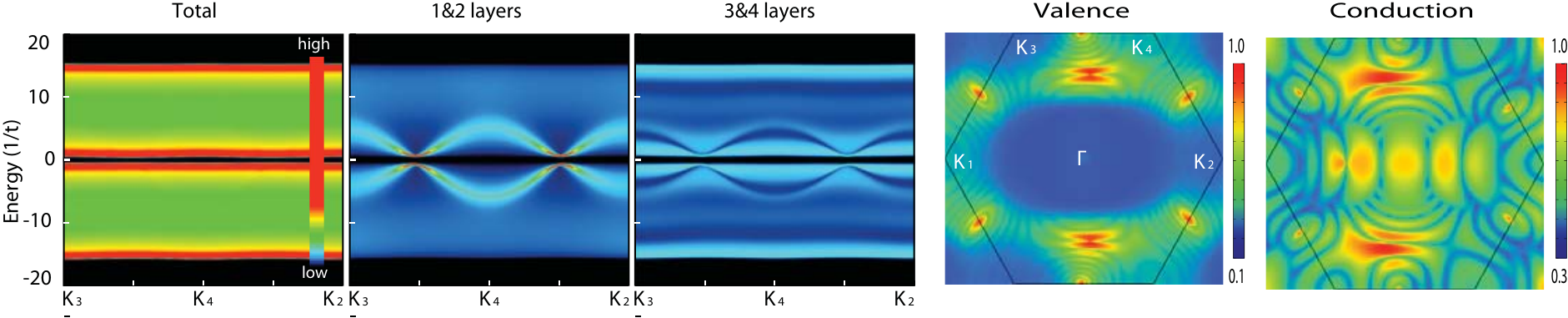}
\caption{{\bf Energy dispersion and distribution of localized states for the
two-dimensional (2D) quantum Hall insulator (QHI) with Chern number two.}
Here the Chern numbers of the conduction and valence bands of the 2D QHI
are -2 and +2, respectively.
Notice that there are several localized bound states in the 2D Brillouin zone
in both the valence and conduction bands of the three-dimensional (3D) normal insulator (NI) substrate as shown in the two panels on the right.
On the left three panels, the dispersion of the coupled heterostructure, which is projected to the layers near the top surface
of the system, is shown along the Brillouin zone boundary.
} \label{fig:C2}
\end{figure*}

\begin{figure*}[t]
\centering
\includegraphics[width=16 cm]{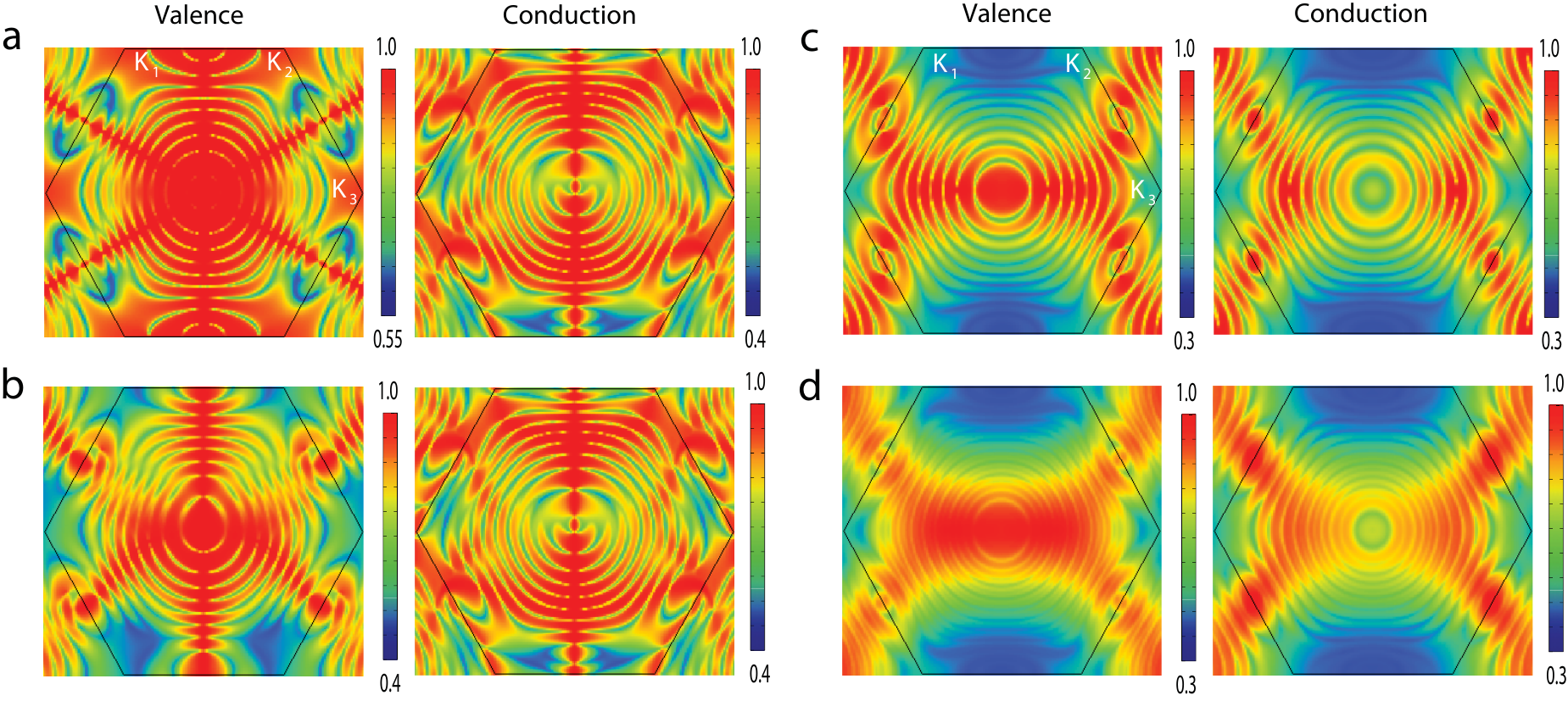}
\caption{{\bf The 2D distribution of localized states of the two-dimensional (2D)
quantum spin Hall insulator (QSHI) bulk bands in three-dimensional (3D) continuum states.}
({\bf a}) and ({\bf b}) corresponds to the case
when the 3D bulk breaks the time-reversal symmetry.
In ({\bf c}) and ({\bf d}), the total system preserves the time-reversal symmetry.
The $z$-component of the total spin ($S_{z}$) is conserved in ({\bf a}) and ({\bf c})
while $S_{z}$ is not conserved in ({\bf b}) and ({\bf d}) due to the Rashba-type
spin mixing term.
In all cases, the two high (low) energy bands of the QSHI are buried
in the conduction (valence) band continuum of the 3D normal insulator (NI).
When the 3D bulk states break the time-reversal symmetry,
there are localized bound states independent of the $S_{z}$ conservation(({\bf a}) and ({\bf b})).
On the other hand, when the total system preserves the time-reversal symmetry,
localized bound states exist when $S_{z}$ is conserved (({\bf c}))
while all 2D states spread along the $z$-direction when $S_{z}$ is not conserved ((\bf d)).
} \label{fig:QSH}
\end{figure*}


\


\


\


\end{document}